\documentclass[aps,showpacs,twocolumn,superscriptaddress,prl,10pt, floatfix]{revtex4-1}
\usepackage{graphicx,color,epsfig,rotate}
\usepackage{amssymb,amsmath,bm,ulem}
\usepackage{dcolumn}  
\usepackage{hyperref} 
\usepackage[usenames,dvipsnames]{xcolor}
\usepackage{subfigure}
\usepackage{float}
\usepackage{wrapfig}
\usepackage{braket}
\usepackage{bm}
\usepackage{graphicx}
\usepackage{color}
\usepackage{dcolumn}
\usepackage{array}

\newcommand{\g} \textbf{}

\begin{document}

\preprint{APS/123-QED}
\title{Diamagnetic Sm$^{3+}$ in the topological Kondo insulator SmB$_6$}  

\author{W. T. Fuhrman}
\email[]{wes@jhu.edu}
\affiliation{Institute for Quantum Matter and Department of Physics and Astronomy, The Johns Hopkins University, Baltimore, Maryland 21218 USA}

\author{J. C. Leiner}
\affiliation{Center for Correlated Electron Systems, Institute for Basic Science (IBS), Seoul 08826, Korea}
\affiliation{Department of Physics and Astronomy, Seoul National University, Seoul 08826, Korea}
\affiliation{Neutron Scattering Division, Oak Ridge National Laboratory, Oak Ridge, TN 37831}

\author{J. W. Freeland}
\email[]{freeland@anl.gov}
\affiliation{Advanced Photon Source, Argonne National Laboratory, Argonne, Illinois 60439, USA}

\author{M. van Veenendaal}
\affiliation{Department of Physics, Northern Illinois University, DeKalb, Illinois 60115, USA}
\affiliation{Advanced Photon Source, Argonne National Laboratory, Argonne, Illinois 60439, USA}

\author{S. M. Koohpayeh}
\affiliation{Institute for Quantum Matter and Department of Physics and Astronomy, The Johns Hopkins University, Baltimore, Maryland 21218 USA}

\author{W. Adam Phelan}
\affiliation{Institute for Quantum Matter and Department of Physics and Astronomy, The Johns Hopkins University, Baltimore, Maryland 21218 USA}
\affiliation{Department of Chemistry, The Johns Hopkins University, Baltimore, Maryland 21218 USA}

\author{T. M. McQueen}
\affiliation{Institute for Quantum Matter and Department of Physics and Astronomy, The Johns Hopkins University, Baltimore, Maryland 21218 USA}
\affiliation{Department of Chemistry, The Johns Hopkins University, Baltimore, Maryland 21218 USA}
\affiliation{Department of Materials Science and Engineering, The Johns Hopkins University, Baltimore, Maryland 21218 USA}

\author{C. Broholm}
\affiliation{Institute for Quantum Matter and Department of Physics and Astronomy, The Johns Hopkins University, Baltimore, Maryland 21218 USA}
\affiliation{Department of Materials Science and Engineering, The Johns Hopkins University, Baltimore, Maryland 21218 USA}
\affiliation{NIST Center for Neutron Research, Gaithersburg, Maryland 20899, USA}

\date{\today}

\begin{abstract}
Samarium hexaboride is a topological Kondo insulator, with metallic surface states manifesting from its insulating band structure. Since the insulating bulk itself is driven by strong correlations, both the bulk and surface host compelling magnetic and electronic phenomena. We employed X-ray absorption spectroscopy (XAS) and X-ray magnetic circular dichroism (XMCD) at the Sm M$_{4,5}$ edges to measure surface and bulk magnetic properties of Sm$^{2+}$ and Sm$^{3+}$ within SmB$_6$. We observed anti-alignment to the applied field of the Sm$^{3+}$ magnetic dipole moment below $T = 75$~K and of the total orbital moment of samarium below 30 K. The induced Sm$^{3+}$ moment at the cleaved surface at 8 K and 6 T implies 1.5\% of the total Sm as magnetized Sm$^{3+}$. The field dependence of the Sm$^{3+}$ XMCD dichorism at 8 K is diamagnetic and approximately linear. The bulk magnetization at 2 K is however driven by Sm$^{2+}$ Van Vleck susceptibility  as well as 1\% paramagnetic impurities with $\mu_{\rm Eff} = 5.2(1)~\mu_{\rm B}$. This indicates diamagnetic Sm$^{3+}$ is compensated within the bulk. The XAS and XMCD spectra are weakly affected by Sm vacancies and carbon doping while XAS is strongly affected by polishing. \end{abstract}

\maketitle

The growing interest and application of topology in condensed matter physics has renewed investigations into SmB$_6$, a cornerstone material of condensed matter and materials science which has now been studied for more than 50 years.\cite{eick1959precise, kasuya1979valence, wolgast2013low} Evidence continues to grow in support of the claim that SmB$_6$ is a topological Kondo insulator, with an insulating bulk at low temperatures and a topologically protected metallic surface.\cite{TKI_theory1, TKI_theory_2, CubicTKI,wolgast2013low,kim2013surface,   eo2018robustness} Unexpected observations related to bulk magnetism have come recently via optical conductivity, muon-spin relaxation, and quantum oscillations which may relate to the surface or bulk.\cite{laurita2016anomalous, biswas2014low, xiang2017bulk, hartstein2017fermi}

In light of the topological aspects of SmB$_6$, great attention has been focused on its surface phenomena.\cite{stern2017surface, kim2013surface, alexandrov2015kondo, zhang2013hybridization, jiang2013observation}  The strongly-correlated nature of the insulating state implies that topological surface states should also be strongly correlated, with potentially exotic implications.\cite{PhysRevLett.114.036401, nikolic2014two, akintola2018lowenergy} Complicating and enriching matters, magnetic impurities have been shown to be common in SmB$_6$, introducing in-gap states and disrupting the surface state.\cite{kim2014topological,fuhrman2017screened, phelan2016chemistry} Samarium vacancies, which are difficult to avoid in floating-zone grown crystals, also produce states in the gap and may contribute to low-temperature thermal transport.\cite{valentine2016breakdown, boulanger2017field} Impurities and defects have been proposed to be related to anomalous low-energy phenomena such as quantum-oscillations.\cite{fuhrman2017screened, shen2018quantum, harrison2018highly} 

In this Letter, surface and bulk sensitive X-ray Magnetic Circular Dichroism (XMCD) measurements are utilized to probe the magnetism related to Sm$^{3+}$ and Sm$^{2+}$ in vacuum cleaved and nominally pure, stoichiometric SmB$_6$ as well as vacuum-cleaved Sm-deficient and carbon-doped samples. The element and valence specific capability of this technique allows  examination of Sm$^{2+}$ and Sm$^{3+}$ moments separately and independently from other contributions to the magnetization. Surprisingly, the data reveal the net magnetization carried by  Sm$^{3+}$ is antialigned to the applied field for temperature ($T$) below 75 K despite positive bulk magnetization. This anomalous diamagnetic component is readily observed at the surface via electronic yield XMCD but it is also indicated by bulk sensitive fluorescence yield XMCD. We relate this observation to paramagnetic impurities and infer that Sm$^{3+}$ interacts antiferromagnetically with larger moment paramagnetic impurities.

\begin{figure}
\includegraphics[totalheight=0.42\textheight,]{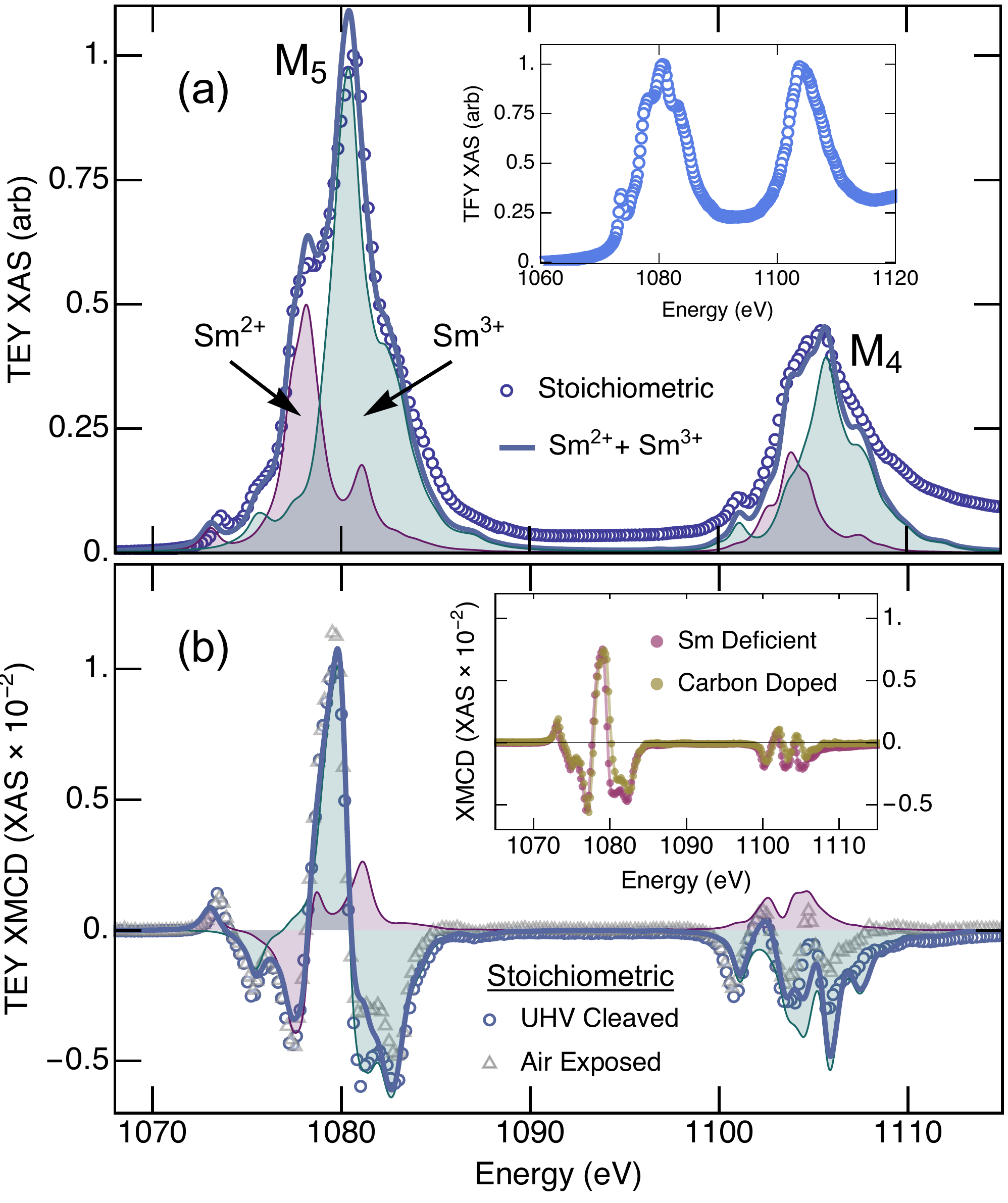}
\caption{ \label{XAS} XAS and XMCD spectra at $T = 8$~K, $\mu_0H=5$~T. Data were normalized by scaling the maximum at the M$_5$ edge (1079 eV). Shaded portions show relative contributions of Sm$^{2+}$ and Sm$^{3+}$. (a) TEY shows the XAS of the surface (approx 2 nm thickness), while TFY spectra show the bulk response (inset). (b) XMCD TEY and linear combination of Sm$^{2+}$ and Sm$^{3+}$ XMCD spectra calculated with Xclaim.\cite{FernandezRodriguez} XMCD was similar for a sample exposed to air (grey triangles). Inset shows XMCD of Sm-deficient and carbon-doped samples. }
 \end{figure}

SmB$_6$ crystals in stoichiometric, carbon-doped, and Sm$_{1-x}$B$_6$ versions were grown using the floating zone (FZ) technique as described by Phelan et al.\cite{phelan2014correlation} Starting materials were polycrystalline SmB$_6$ rods (Testbourne Ltd, 99.9\%). Previous elemental analysis indicated rare-earth and alkaline earth impurities at the 10$^3$ ppm scale. These impurities form stable hexaborides with similar lattice parameters to SmB$_6$ and thus predominantly occupy the Sm-site; summing up their concentrations indicates approximately 2\% (1\% magnetic with weighted average moment $\mu_{\rm avg}$ = 5$\mu_{\rm B}$) impurities per formula unit.\cite{phelan2016chemistry} 

The XAS and XMCD measurements were conducted at beam line 4-ID-C of the Advanced Photon Source located at Argonne National Laboratory. SmB$_6$ crystals were notched to facilitate (100) cleavage. Crystals were cleaved after placement in the vacuum chamber ($8\times 10^{-9}$ Torr) for measurement. Surface and bulk sensitive XAS and XMCD spectra were collected simultaneously using total electron yield (TEY) and total fluorescence yield (TFY) respectively with circularly polarized x-rays in a near normal (80$^{\circ}$) configuration. The applied field was along the beam direction and it defines the positive $\hat{z}$ direction. The TEY mode probes approximately the first 2 nm of the SmB$_6$ surface, while TFY is bulk-sensitive. The XMCD spectra were obtained point-by-point by subtracting right from left circular polarized XAS data. Measurements  were taken for both positive and negative applied field directions and then we take a difference of these two spectra XMCD=$\frac{1}{2}$(XMCD($H_z>0$)$-$XMCD($H_z<0$)) to eliminate polarization dependent systematic errors. The stoichiometric sample central to this study was cleaved more than 24 hours before measurement, sufficient time for complete surface reconstruction.\cite{zabolotnyy2018chemical} 

The isotropic and dichroic x-ray absorption spectra were calculated using Xclaim\cite{FernandezRodriguez} in the atomic limit \cite{Thole,Goedkoop}, which is appropriate for rare earth 4$f$ electrons. The Hamiltonian includes spin-orbit interaction in the $3d$ and $4f$ orbitals and Coulomb interactions in the $4f$ shell and between the $4f$ shell and the $3d$ core hole. Parameters were obtained in the Hartree-Fock limit and the values for the Coulomb interaction were scaled down to 80\% to account for screening effects. The calculated spectra are consistent with pure, divalent, and trivalent Sm compounds. Fits of relative Sm$^{2+}$ and Sm$^{3+}$ contributions allow for a small shift in energy ($<$1 eV) with fixed relative energy profiles. 

\begin{table}
  \begin{tabular*}{0.485\textwidth}{@{\extracolsep{\fill} } |l|l|l|}
  \hline
& Sm$^{3+}$ & Sm$^{2+}$ \\ [0.5ex] 
  \hline
 $\left< L_z \right>$  ($\hbar$) & -0.16(3) & 0.10(2)  \\ 
 $\left< S_z \right>$  ($\hbar$) & 0.06(1) & -0.10(2) \\ 
 \hline
 $\left< J_z \right>$ ($\hbar$)& -0.10(3) & 0.00(2) \\
  \hline
$\left<  L_z \right>$/ $\left< S_z \right>$ & -2.66 & -1.00 \\
 \hline
\end{tabular*}

 \caption{Sm$^{3+}$ and Sm$^{2+}$ contributions to the z-component of the orbital and spin magnetic moments obtained from fits to the TEY XMCD spectra of stoichiometric SmB$_6$.
 }
\label{Table_LZ}
\end{table}

The XAS near the M$_5$ (1080 eV) and M$_4$ (1105 eV) absorption edges (Fig.~\ref{XAS}(a)) shows distinct peaks from Sm$^{2+}$($4f^6$) and Sm$^{3+}$($4f^5$) in both the TEY and TFY channels. At the M edges, the bulk sensitive TFY XMCD signal is weak and distorted by self-absorption effects, and so we proceed first with analysis of the surface sensitive TEY XMCD.\cite{pompa1997experimental, nakazawa1998theory, van2014x}  In field at low temperatures, the presence of both divalent and trivalent Sm is clearly visible in the pre-edge region of M$_5$ where their dichroism features are opposite. Additionally, the main line of Sm$^{2+}$ causes a significant negative dichroic feature around 1077 eV that is not present in the dichroic spectrum of Sm$^{3+}$.  The contributions of Sm$^{3+}$ and Sm$^{2+}$ to the TEY XMCD spectrum are shown in Table 1. This dichroic spectrum is evidence of magnetizable moments at the surface of SmB$_6$. Because this response is observed in vacuum cleaved samples, it cannot be attributed to the formation of surface oxides. The spectra of a sample exposed to air for several days showed little change in the dichroic spectra. The TEY XMCD in field is similar for carbon-doped, and Sm-deficient samples (inset of Fig.~\ref{XAS}(b)), with integrated mean squared XMCD ($\int_{M_{4,5}}\sqrt[]{{\rm XMCD}^2}$) of 0.017 (pure), 0.014 (Sm deficient), 0.013 (carbon-doped).

\begin{figure}
\includegraphics[totalheight=0.48\textheight]{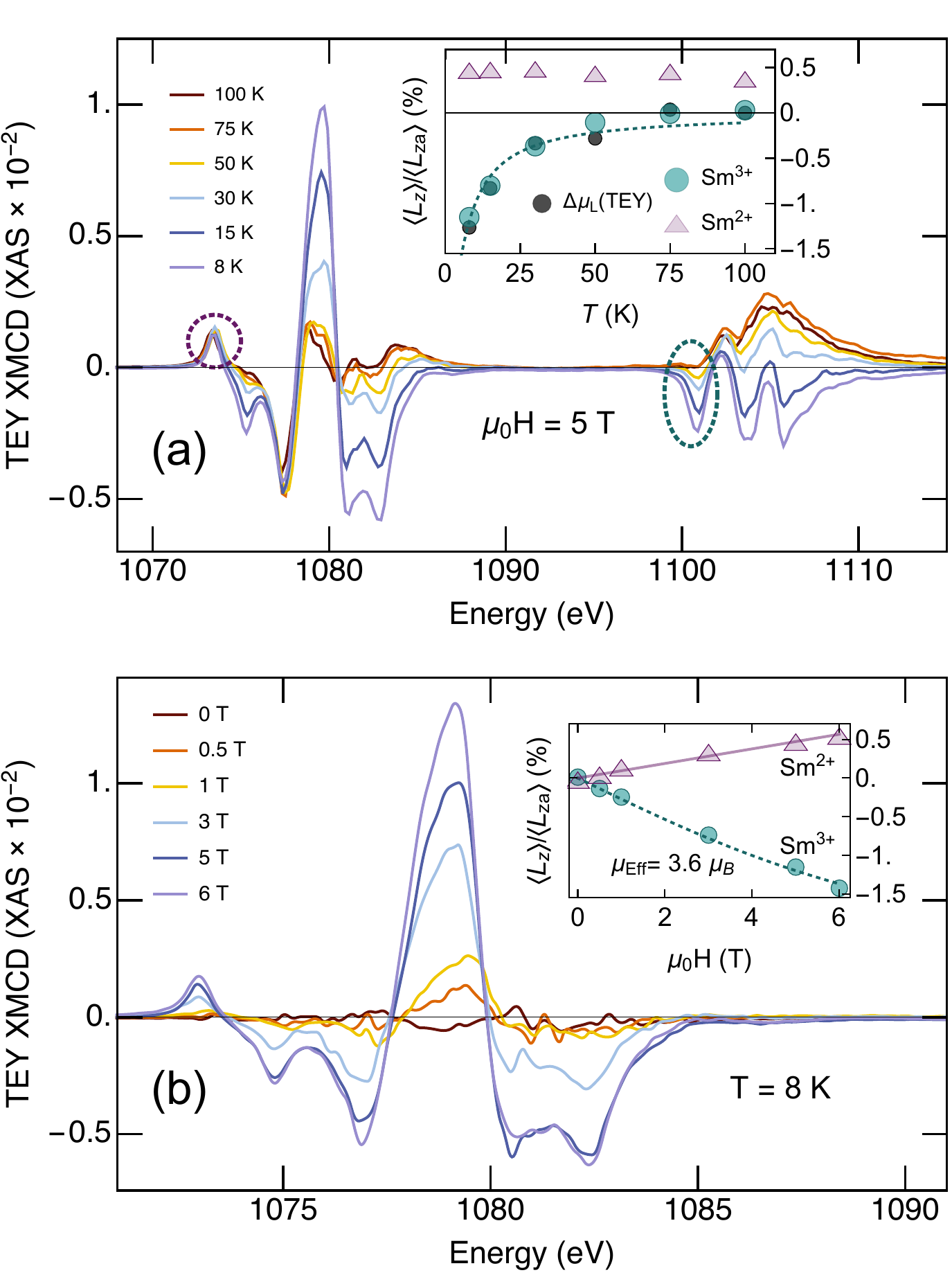}
\caption{\label{TempField} 
TFY(surface) XMCD temperature and magnetic field dependence. (a) XMCD temperature dependence.  Circled energies in the main panel indicate TEY XMCD spectra dominated by a single valence (1073.5 eV for Sm$^{2+}$ and 1100.5 eV for Sm$^{3+}$) Inset shows temperature dependence of the fitted Sm$^{2+}$ and Sm$^{3+}$ XMCD amplitudes and integrated $\Delta$XMCD relative to 100 K ($\propto \Delta \mu_{\rm L}$). (b) Magnetic field response of the M$_5$ edge TEY XMCD at 8 K. Inset shows the contributions from Sm$^{2+}$  and Sm$^{3+}$. In the insets, the dotted lines show a Langevin fit ($\mu_{\rm Eff} = 3.6(9) \mu_{\rm B}$, concentration 2.7(5)\%) of the combined temperature dependence below 75 K and field dependence at 8 K}
 \end{figure}

At higher temperatures, the TEY XMCD is predominantly associated with Sm$^{2+}$ (Fig. \ref{TempField}(a)). This contribution is evidenced by the Sm$^{2+}$ pre-edge M$_5$ peak at 1073.5 eV, which further shows little temperature dependence. While Van Vleck type $J=0$ Sm$^{2+}$ paramagnetism has only weak temperature dependence, free $J=5/2$ Sm$^{3+}$ carries a magnetic moment which should give rise to a Curie term $\propto 1/T$ in the corresponding magnetic susceptibility. Upon cooling below 75 K a substantial feature at M$_5$ develops along with a weaker M$_4$ structure. The predominantly Sm$^{3+}$ leading edge of M$_4$ (1100.5 eV) and fitted Sm$^{3+}$ contribution show these dichroic features are associated with magnetized Sm$^{3+}$ with net diamagnetism as the material becomes more insulating. 

In addition to the fitting described above, sum rule analysis directly provides the Sm orbital moment through integration of the XMCD spectra over both the M$_4$ and M$_5$ edges (Fig.2(a) inset).\cite{carra1993x} Given the weak temperature dependence of the Van Vleck Sm$^{2+}$ component, the total orbital moment extracted from the TEY XMCD through sum rule analysis is expected to follow the fitted Sm$^{3+}$ component, offset by a constant. At high temperatures, the total orbital moment is positive, changing sign as temperature is reduced below 30 K. This change in sign to a negative total orbital moment at low-temperatures is model-independent evidence of a net diamagnetic orbital magnetic moment carried by Sm at low~$T$. Subtracting off the high-$T$~(100~K) Sm$^{2+}$ component, we can compare the change in orbital moment (related to Sm$^{3+}$) to the expected Hund's rule value of $\left<L_{z_a}\right>$ = 5, finding $\Delta\left<L_z\right>/ \left<L_{z_a}\right>$ = 1.5\% of the total Sm as magnetized Sm$^{3+}$ at 8 K and 6 T. For reference, at 8 K and 6 T   small-moment Sm$^{3+}$ ($\mu_{\rm Eff}$ = 0.85 $\mu_B$) yields 14\% of its saturated moment while large moment impurities ($\mu_{\rm Eff}\approx 5$) should be magnetized to 63\% of their saturated moment.

The temperature and field dependence of the change in TEY (surface) XMCD, $\Delta\left<L_z\right>/ \left<L_za\right>$, can be fit by a negative Langevin function, $L(x) = c ({\rm coth}(x)-1/x)$, where $c$ is the concentration and $x$ is the product of effective moment, field, and inverse temperature, $x = \mu_{\rm Eff}~\mu_oH/(k_B T)$. This fit yields $\mu_{\rm Eff} = 3.6(9)\mu_{\rm B}$ with a concentration of 2.7(5)\% of the total population of Sm at saturation. This moment is larger than the Sm$^{3+}$ moment, but close to the weighted average impurity moment. The implied concentration is also similar to the impurity concentration. In zero field (Fig.~\ref{TempField}(b)), the TEY XMCD shows no evidence of magnetization beyond the experimental detection limit ($<2$\% of the 5 T response at the M$_5$ peak, 1079.5 eV), an indication against surface ferromagnetism at 8 K.  However, these magnetic components may have a magnetically ordered phase at sufficiently low temperatures, as suggested by hysteretic magnetotransport.\cite{wolgast2015magnetotransport}

\begin{figure}
\includegraphics[totalheight=0.43\textheight]{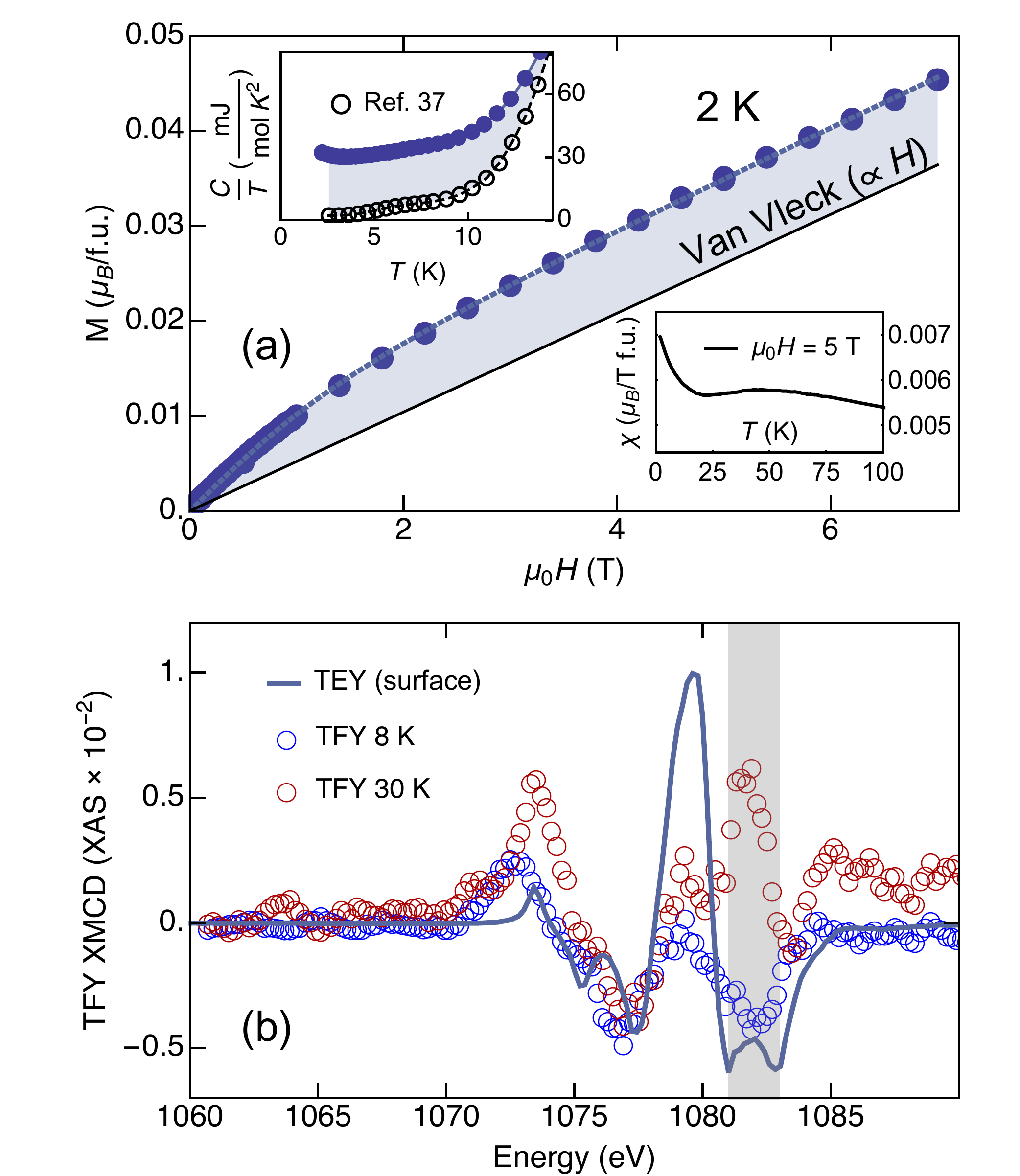}
\caption{\label{Mag} Bulk properties of nominally pure SmB$_6$ sample. (a) The magnetization data is fit by a Van Vleck contribution (solid black line) and a paramagnetic impurity contribution (shaded) of 1\% impurites with $\mu_{\rm Eff}$~=~5.2~$\mu_{\rm B}$. Insets show susceptibility taken at 5 T and heat capacity with comparison to the previously published heat capacity of a high-purity sample.\cite{orendavc2017isosbestic} We attribute shaded portions to impurities. Sample data also appears in the supplementary information of Ref.~\cite{phelan2016chemistry}, without fitting. (b) TFY XMCD (bulk). At 8 K, a negative dichroic feature develops from 1081 eV - 1083 eV as for TEY. The temperature dependence counter-indicates solely Sm$^{2+}$. The temperature dependence of the integrated TFY XMCD is shown in the inset of Fig.2(a).}
 \end{figure}

The bulk-sensitive TFY XMCD also indicates dichroic features within the bulk of SmB$_6$~(Fig.4 inset). If the bulk XMCD signal were entirely Sm$^{2+}$ in origin, it would be expected to carry the weak temperature dependence seen of Sm$^{2+}$ in TEY. However, upon cooling, the trailing edge of M$_5$ develops a dichroic feature which mirrors that of the surface (1081 eV-1083 eV). The TEY and TFY dichroic features are similar in magnitude, and the change in integrated TFY XMCD decreases with lowering temperature. This suggests diamagnetism of Sm$^{3+}$ within the bulk as well as at the surface of SmB$_6$. 

To contextualize the diamagnetic XMCD Sm$^{3+}$ with the net magnetic properties of  bulk SmB$_6$, we investigated the magnetization and susceptibility of the stoichiometric sample (Fig. 3(a)), reported previously without analysis.\cite{phelan2014correlation} A flattening of the susceptibility (Fig. 3(a) inset) occurs at 60 K, forming a broad hump before an eventual upturn at low $T$. The rounded maximum suggests short range antiferromagnetic correlations. The low $T$ upturn is variable across samples of SmB$_6$ and can be attributed to a Curie-like susceptibility of weakly interacting magnetic impurities. At low-temperatures, M(H) is well fit by the sum of a linear component (M = 0.0052~$\mu_{\rm B} T^{-1}$ f.u.$^{-1}$) associated with Van Vleck magnetism and a Langevin function of 1\% magnetic impurities with an effective moment 5.2(1)~$\mu_{\rm B}$ (Fig. 3(a)). Such fits have been shown to be effective over wide ranges of impurity concentrations, fields, and temperatures in SmB$_6$.\cite{fuhrman2017screened} The overall positive moment seen in bulk magnetization measurements indicates the predominant contribution to the low $T$ uniform magnetization is not the negative-moment Sm$^{3+}$ magnetism seen by XMCD. However, Sm$^{3+}$ coupled antiferromagnetically to larger moment impurities would appear diamagnetic when observed independently.

The observed bulk magnetization is also consistent with screening of magnetic impurities inferred from elemental analysis. While the XAS edges probed here limit sensitivity to Sm 4$f$ electrons, previously described moment-screening in Gd-doped SmB$_6$ provides a basis for comparison.\cite{fuhrman2017screened}. Assuming a similar effect
, the expected moment screening of the bulk magnetization for 1\% magnetic impurities (known to be present through elemental analysis\cite{phelan2016chemistry}) would be 10\% (.05$\mu_{\rm B}$), similar to the inferred bulk Sm3+ diamagnetism of order 1\% of $\mu_{\rm Sm} = 0.85 \mu_{\rm B}$ Sm$^{3+}$ ($.0085\mu_{\rm B}$). 
The enhanced low-temperature heat capacity seen in our stoichiometric sample relative to a high-purity sample (Fig.3(a) inset) is also consistent with the enhanced heat capacity associated with impurities and moment screening. High-quality starting materials yield SmB$_6$ samples with more than an order of magnitude smaller heat capacity  at 2~K.\cite{orendavc2017isosbestic}  


To determine the effect of surface preparation, we measured the polished surface of a stoichiometric sample from the same portion of the floating zone grown sample for comparison with our in-situ vacuum cleaved sample. The ratio of $\left< L_z \right>$ to its atomic (local) value ($\left< L_{za} \right>$) for the stoichiometric cleaved and polished samples differ between samples despite identical starting materials and presumed stoichiometric similarity. While the Sm$^{3+}$ signal is similar between samples ($\left< L_{z_{\rm cleaved \vphantom{p}}}^{3+}\right>$/$\left< L_{z_{\rm polished}}^{3+}\right> = 0.99)$, the Sm$^{2+}$ signal changes by nearly a factor of two ($\left< L_{z_{\rm cleaved\vphantom{p}}}^{2+}\right>$/$\left< L_{z_{\rm polished}}^{2+} \right>$ = 1.8). 

Surprisingly, polishing has a more substantial effect than carbon-doping and Sm vacancies on the average Sm valence in our samples. Our stoichiometric cleaved samples have valence 2.64(3) while polished have valence 2.77(3). Cleaved samples of Sm-deficient and carbon-doped samples have valence 2.64(3) and 2.60(3), respectively. Given that SmB$_6$ is close to a trivial insulator phase dictated by valence, caution is warranted in preparing materials and surfaces for which topological properties are measured.\cite{alexandrov2013cubic} 

We have observed a diamagnetic response for magnetizable Sm$^{3+}$ below 75 K in SmB$_6$ via XMCD. 
The moment is anti-aligned with the applied field and paramagnetic-like in field at 8 K. The XMCD signal is weakly sensitive to carbon doping and Sm-deficiencies, indicating the negative Sm$^{3+}$ moment is intrinsic or related to shared impurities, shown to be at the level of 2\% (1\% magnetic). The bulk magnetization distinctly requires that the observed negative Sm$^{3+}$ moment is either absent or overwhelmed within the bulk. The bulk-sensitive TFY XMCD, though noisy, is consistent with a negative Sm$^{3+}$ moment. If the observed XMCD is intrinsic and present within the bulk, this implies the Kondo singlet ground state is modified by magnetic field despite previous magnetization measurements on higher-purity samples showing almost exclusively Van Vleck magnetization to at least 60 T.\cite{tan2015unconventional} An exotic form of diamagnetism has been proposed at very low fields for SmB$_6$.\cite{erten2017skyrme} However, given the known impact of even modest impurity concentrations on the low-energy physics within SmB$_6$  and the Curie-like temperature dependence of the Sm$^{3+}$ XMCD, a natural explanation for our observations is that the magnetizable Sm$^{3+}$ in our SmB$_6$ samples anti-aligns to the applied field as a consequence of strong antiferromagnetic coupling to larger moment impurities. In this way the diamagnetic response that we detect for Sm$^{3+}$ is associated with bulk compensated paramagnetism. 

This project was supported by UT-Battelle LDRD 3211-2440. The work at IQM was supported by the US Department of Energy, office of Basic Energy Sciences, Division of Material Sciences and Engineering under grant DE-FG02-08ER46544. W.T.F. is grateful to the ARCS foundation, Lockheed Martin, and KPMG for the partial support of this work. MvV was supported by the U. S. Department of Energy (DOE), Office of Basic Energy Sciences, Division of Materials Sciences and Engineering under Award No. DE-FG02-03ER46097. Work at Argonne National Laboratory was supported by the U. S. DOE, Office of Science, Office of Basic Energy Sciences, under contract No. DE-AC02-06CH11357.

\bibliography{SmB6_ref} 

\end{document}